\documentstyle[amssymb,twocolumn,aps,pra,psfig]{revtex}

\begin{document}
\title{Quantum theory of fluctuations in a cold damped accelerometer}
\author{Francesca Grassia $^a$ \footnote{grassia@spectro.jussieu.fr}, 
Jean-Michel Courty $^a$ \footnote{courty@spectro.jussieu.fr}, 
Serge Reynaud $^a$ \footnote{reynaud@spectro.jussieu.fr}
and Pierre Touboul $^b$ \footnote{touboul@onera.fr}}
\address{(a) Laboratoire Kastler Brossel \thanks{Laboratoire de 
l'Universit\'{e} Pierre et Marie Curie, de l'Ecole
Normale Sup\'{e}rieure et du Centre National de la Recherche
Scientifique.}, UPMC case 74, \\
4 place Jussieu, F\ 75252 Paris Cedex 05 \\
(b) D\'{e}partement de Mesures Physiques, ONERA, \\
29 Av. de la division Leclerc, BP72, F\ 92322 Chatillon Cedex}
\date{July 1999}
\maketitle

\begin{abstract}
We present a quantum network approach to real high sensitivity measurements.
Thermal and quantum fluctuations due to active as well as passive elements
are taken into account. The method is applied to the analysis of the
capacitive accelerometer using the cold damping technique, developed for
fundamental physics in space by ONERA and the ultimate limits of this
instrument are discussed. It is confirmed in this quantum analysis that the
cold damping technique allows one to control efficiently the test mass
motion without degrading the noise level.

\end{abstract}

\section{Introduction}

When discussing ultimate limits in ultrasensitive measurements, we have to
take into account fundamental fluctuation processes as well as a realistic
description of the measurement device. This requires to treat in the same
theoretical framework a number of problems which are often tackled by
different approaches. Real measurements always have a finite time
resolution, that is also a characteristic frequency bandwidth, as well as a
finite duration. The measurement is never infinitely precise and
fluctuations are superimposed to the signal. Ultrasensitive measurement
devices often make use of active systems either for amplifying the signal to
a readable level or to make the system work around its optimal operating
point with the help of feedback loops. Feedback loops can also be used to
modify the natural frequency response and, in particular, to perform an
optimal damping of moving elements.

The aim of the present paper is to develop an approach of ultrasensitive
measurements taking into account these various problems. In particular, we
want to treat thermal as well as quantum fluctuations for systems containing
active as well as dissipative elements. The approach will be illustrated by
analyzing the sensitivity of a cold damped capacitive accelerometer
developed for fundamental physics applications in space \cite
{Bernard91,Touboul92,Willemenot97}. In this measurement system, feedback
loops are used to keep the proof mass perfectly centered in the
accelerometer cage and to damp its motion without adding the thermal
fluctuations which would necessarily accompany a passive damping. With this
technique, fluctuations are reduced to an effective temperature well below
the operating temperature \cite{Milatz53}. The cold damping technique is
known to be compatible with very high sensitivities of the measurement \cite
{Vitale}. However the question of ultimate sensitivities compatible with the
existence of quantum fluctuations remains open. This question is important
not only for a better understanding of the instrument but also for the long
term purpose of an improvement of its performances. For earth based
detection of gravitational waves, highly effective motion isolation and
feedback controlled noise reduction is developped \cite{Newell97,Blair}. The cryogenic
accelerometers planned for future space mission such as LISA will require
very low noise levels and they use cryogenic techniques.

Relations between fluctuations and dissipation have been first discovered by
Einstein which studied the viscous damping of mechanical systems \cite
{Einstein05}. Another important application was the study of Johnson-Nyquist
noise in resistive electrical elements \cite{Nyquist28}. These general
thermodynamical relations were widely studied in the framework of linear
response theory \cite{Kubo66,Landau}. In the limit of a null temperature,
they reproduce quantum fluctuations required by Heisenberg inequalities \cite
{Callen51}. Important progress have been made during the last two decades
towards a better control of the effect of quantum fluctuations on
ultrasensitive measurements \cite{Yamamoto90,Reynaud92}. It has been shown
that it was possible to bypass the limitations usually associated with
quantum noise by using back action evading measurements or quantum non
demolition techniques \cite{Braginsky92,Grangier92,Ricci96,Bocko96}.
Fluctuations associated with amplification were also extensively studied 
\cite{Heffner62,Haus62,Gordon63}; they determine the ultimate performance of
linear amplifiers \cite{Caves82,Loudon84} and they may be used to reduce
inloop quantum fluctuations with feedback \cite{Yamamoto 86,Liebman93}.

In the present paper we will study this kind of measurement systems by using
a systematic approach which may be termed as ``quantum network theory''.
Initially designed as a quantum extension of the classical theory of
electrical networks \cite{Meixner63}, this theory was mainly developed
through applications to optical systems \cite{Yurke84,Gardiner88}. It can be
viewed as a generalization of the linear response theory \cite{Courty92} and
is also fruitful for analysing non-ideal quantum measurements with active
elements \cite{Francesca98} as soon as a quantum theory of ideal operational
amplifier is available \cite{Courty99}. The main features of this approach
are recalled in section 2.

Here, this theory will be illustrated by a study of an electromechanical
measurement system comprising active and dissipative components coupled to a
capacitive position sensor. The parametric nature of the electromechanical
coupling allows the use of a frequency transfer technique in order to
eliminate the influence of the $1/f$ noise in the electric part of the
device. Active elements are used for preamplifying the position sensing
signal to a readable level as well as for controlling the mechanical motion
through a feedback loop. The main features of the cold damped capacitive
accelerometer are described in section 3. Then the capacitive sensor is
analysed in absence of servo control in section 4 and in presence of servo
control in section 5. These results are used in section 6 to evaluate the
ultimate sensitivity of the measurement system, which is found to be
essentially determined by the free mechanical impedance of the proof mass
and the ratio of the frequencies involved in the frequency transfer
performed by the transducer.

\section{Noise in electromechanical systems}

In this section we present the basic elements of the quantum network
approach. Quantum and thermal fluctuations in dissipative and active systems
are all described in terms of quantum fields. All the descriptions are given
in the frequency domain and the convention of quantum mechanics is used for
the Fourier transform. The electronics convention may be recovered by
substituting $j$ to $-i$.

In a quantum network approach, the various fluctuations entering the system,
either by dissipative or by active elements, are described by input fields
in noise lines coupled to a reactive network (see Figure 1).

\begin{figure}[htb]
\caption{Representation of an electrical circuit as a quantum network. The
central box is a reactive multipole which connects noise lines corresponding
to the fluctuations entering the system, either by dissipative or by active
elements. For example, the upper left port $p$ with voltage $U_{p}$ and
current $I_{p}$ is connected to a line of impedance $R_{p}$ with inward and
outward fields $p^{{\rm in}}$ and $p^{{\rm out}}$.}
\label{fig1}
\end{figure}

In particular, a resistance $R_{p}$ is modeled as a semi-infinite coaxial
line $p$ with characteristic impedance $R_{p}$. The voltage $U_{p}$ and
current $I_{p}$ associated with the resistance are the inward and outward
fields $p^{{\rm in}}$ and $p^{{\rm out}}$ evaluated at the end of this line 
\begin{eqnarray}
I_{p} &=&\sqrt{\frac{\hbar \left| \omega \right| }{2R_{p}}}\left( p^{{\rm out%
}}-p^{{\rm in}}\right)  \nonumber \\
U_{p} &=&\sqrt{\frac{\hbar \left| \omega \right| R_{p}}{2}}\left( p^{{\rm out%
}}+p^{{\rm in}}\right)  \label{defnyquist}
\end{eqnarray}
These equations may be written equivalently 
\begin{eqnarray}
U_{p} &=&R_{p}I_{p}+\sqrt{2\hbar \left| \omega \right| R_{p}}p^{{\rm in}} 
\nonumber \\
p^{{\rm out}} &=&\sqrt{\frac{2}{\hbar \left| \omega \right| R_{p}}}U_{p}-p^{%
{\rm in}}  \label{resistance}
\end{eqnarray}
The first equation in (\ref{resistance}) is the standard current-voltage
relation for a resistance with the Johnson-Nyquist noise described as the
input fields $p^{{\rm in}}$ going to the end of the line. The second
equation gives the output fields $p^{{\rm out}}$ emitted back to the line.\
In the following, these fields are used either to feed other elements of the
system or to perform a measurement by extracting information from the system
of interest through a line considered as the detection channel.

Input fields $p^{{\rm in}}$ are described as free fields in a
two-dimensional quantum field theory. They obey the standard commutation
relation of such a theory 
\begin{equation}
\left[ p^{{\rm in}}\left[ \omega \right] ,p^{{\rm in}}\left[ \omega ^{\prime
}\right] \right] =2\pi \ \delta \left( \omega +\omega ^{\prime }\right) \
\varepsilon \left( \omega \right)  \label{freecommut}
\end{equation}
where $\varepsilon \left( \omega \right) $ denotes the sign of the frequency 
$\omega $. This relation just means that the positive and negative frequency
components correspond respectively to the annihilation and creation
operators of quantum field theory. Input fields corresponding to different
lines commute with each other. For simplicity, the fields incoming through
the various ports are supposed to be uncorrelated with each other. The
interaction with non linear reactive elements are linearized around the
working point of the system. With the whole network is then associated a
scattering $S$ matrix, also called repartition matrix, describing the
tranformation from the input fields to the output ones. The output fields $%
p^{{\rm out}}$ are also free fields which obey the same commutation
relations (\ref{freecommut}) as the input ones. Hence, the $S$ matrix must
be unitary in order to preserve the field commutation relations.

Input fluctuations are characterized by a noise spectrum $\sigma _{pp}^{{\rm %
in}}$ with its well-known expression for a thermal equilibrium at a
temperature $T_{p}$ 
\begin{eqnarray}
\left\langle p^{{\rm in}}\left[ \omega \right] \cdot p^{{\rm in}}\left[
\omega ^{\prime }\right] \right\rangle &=&2\pi \ \delta \left( \omega
+\omega ^{\prime }\right) \ \sigma _{pp}^{{\rm in}}\left[ \omega \right] 
\nonumber \\
\sigma _{pp}^{{\rm in}}\left[ \omega \right] &=&\frac{1}{2}\coth \frac{\hbar
\left| \omega \right| }{2k_{B}T_{p}}  \label{thermal}
\end{eqnarray}
The symbol `$\cdot $' denotes a symmetrized product for quantum operators
and $k_{B}$ is the Boltzmann constant. The energy per mode will be denoted
in the following as an effective temperature 
\begin{equation}
k_{B}\Theta _{p}=\hbar \left| \omega \right| \sigma _{pp}^{{\rm in}}=\frac{%
\hbar \left| \omega \right| }{2}\coth \frac{\hbar \left| \omega \right| }{%
2k_{B}T_{p}}  \label{effectemp}
\end{equation}
This effective temperature $k_{B}\Theta _{p}$ reproduces the zero point
energy $\frac{\hbar \left| \omega \right| }{2}$ at the limit of zero
temperature and the classical result $k_{B}T_{p}$ at the high temperature
limit. The output fields are also characterized by noise spectra $\sigma
_{pp}^{{\rm out}}$ which are different from those associated with input
fields, due to the interaction with the system. In fact the analysis of the
measurement sensitivity essentially consists in an evaluation of these
functions.

In the capacitive sensor used in the accelerometer, a frequency
transposition technique is used to reduce the $1/f$ electrical noise. The
mechanical signal at frequency $\Omega $ is imprinted on the sidebands $%
\omega _{t}\pm \Omega $ of an electrical carrier oscillating at frequency $%
\omega _{t}$. Such a signal is described by quadrature components 
\begin{eqnarray}
p_{1}\left[ \Omega \right] &=&p\left[ \omega _{t}+\Omega \right] +p\left[
-\omega _{t}+\Omega \right]  \nonumber \\
p_{2}\left[ \Omega \right] &=&\frac{p\left[ \omega _{t}+\Omega \right] +p%
\left[ -\omega _{t}+\Omega \right] }{i}  \label{defquadrat}
\end{eqnarray}
Assuming that $\omega _{t}\gg \Omega $, the noise spectra of these
quadratures is given by 
\begin{equation}
\sigma _{p_{1}p_{1}}^{{\rm in}}=\sigma _{p_{2}p_{2}}^{{\rm in}}=\frac{%
2k_{B}\Theta _{p}}{\hbar \omega _{t}}
\end{equation}
where $\Theta _{p}$ is evaluated from (\ref{effectemp}) for a frequency
equal to $\omega _{t}$.

The previous discussion of electrical elements is easily extended to include
mechanical elements. A mass damped by a viscous force is described by
equations similar to (\ref{defnyquist}) 
\begin{eqnarray}
V_{m} &=&\sqrt{\frac{\hbar \left| \Omega \right| }{2H_{m}}}\left( m^{{\rm out%
}}-m^{{\rm in}}\right)  \nonumber \\
F_{m} &=&\sqrt{\frac{\hbar \left| \Omega \right| H_{m}}{2}}\left( m^{{\rm out%
}}+m^{{\rm in}}\right)  \label{defLangevin}
\end{eqnarray}
or equivalently 
\begin{eqnarray}
F_{m} &=&H_{m}V+\sqrt{2\hbar \left| \Omega \right| H_{m}}m^{{\rm in}} 
\nonumber \\
m^{{\rm out}} &=&\sqrt{\frac{2}{\hbar \left| \Omega \right| H_{m}}}F_{m}-m^{%
{\rm in}}
\end{eqnarray}
In these equations, $H_{m}$ is the friction coefficient, $V_{m}$ the
velocity of the mass, $F_{m}$ the force acting on the mass, $\Omega $ the
mechanical frequency and $m^{{\rm in}}$ and $m^{{\rm out}}$ are input and
output quantum fields in an equivalent mechanical line $m.$ In particular,
the fluctuating Langevin force is proportional to the input fluctuations $m^{%
{\rm in}}$. The free fields $m^{{\rm in}}$ and $m^{{\rm out}}$ obey the same
commutation relation (\ref{freecommut}) as for electrical lines and an
effective temperature is defined as in (\ref{effectemp}) 
\begin{equation}
k_{B}\Theta _{m}=\hbar \left| \Omega \right| \sigma _{mm}^{{\rm in}}=\frac{%
\hbar \left| \Omega \right| }{2}\coth \frac{\hbar \left| \Omega \right| }{%
2k_{B}T_{m}}  \label{effectempM}
\end{equation}

The description of fluctuations in active elements requires further
developments. In the present paper, attention is restricted to active
elements built on ideal operational amplifiers working in the limits of an
infinite input impedance, a null output impedance and an infinite gain. Such
an amplifier is described as a quantum network connected to the left (input)
port, the right (output) port and two lines needed to describe these noise
generators associated with the amplifier \cite{Courty99}. 
\begin{figure}[htb]
\caption{Representation of the ideal operational amplifier as a quantum
network with a left (input) port $l$ and a right (output) port $r$. The
input and output impedances are respectively infinite and null. The
amplifier works in the limit of infinite gain with a reactive feedback $Z_{f}
$. The voltage and current noises of the amplifier are modeled as input
fields in the two noise lines $a$ and $b$.}
\label{fig2}
\end{figure}
The equations of the amplifier, schematized on Figure 2, are read as 
\begin{eqnarray}
U_{l}\left[ \omega \right] &=&U_{r}\left[ \omega \right] +Z_{f}I_{f}\left[
\omega \right]  \nonumber \\
&=&\sqrt{2\hbar \left| \omega \right| R_{a}}\left( a^{{\rm in}}\left[ \omega %
\right] -b^{{\rm in}}\left[ -\omega \right] \right)  \nonumber \\
I_{l}\left[ \omega \right] +I_{f}\left[ \omega \right] &=&\sqrt{\frac{2\hbar
\left| \omega \right| }{R_{a}}}\left( a^{{\rm in}}\left[ \omega \right] +b^{%
{\rm in}}\left[ -\omega \right] \right)  \label{amplifier}
\end{eqnarray}
$U_{l}$ and $U_{r}$ are the voltages at the left and right ports, $I_{l}$
the current at the left port, $I_{f}$ the current across the reactive
impedance $Z_{f}$ ($%
\mathop{\rm Re}%
Z_{f}=0$) used to adjust the transimpedance gain of the amplifier. The
voltage noise and current noise associated with the amplification are
described by two fields $a^{{\rm in}}$ and $b^{{\rm in}}$ which verify the
free field commutation relation (\ref{freecommut}). The field $b^{{\rm in}}$
appears in the equation after a conjugation which interchanges annihilation
and creation operators. The presence of such a conjugation, already known
for linear amplifiers \cite{Caves82,Loudon84}, plays an important role when
commutators are evaluated. It can be forgotten when symmetrized correlation
functions are computed and will be considered as implicit in forthcoming
equations. In (\ref{amplifier}), the impedance $R_{a}$, which characterizes
the amplifier noise, is derived from the ratio of the voltage and current
noises 
\begin{equation}
R_{a}=\sqrt{\frac{\sigma _{UU}}{\sigma _{II}}}
\end{equation}
These fluctuations have been assumed to be phase-insensitive, {\it i.e. }to
be the same for any field quadrature. Although these assumptions are not
mandatory for the forthcoming analysis, the impedance $R_{a}$ is considered
as constant over the spectral domain of interest and the effective
temperature $\Theta _{b}$ is taken equal to $\Theta _{a}$.

\section{General description of the accelerometer}

The capacitive accelerometer operation is presented on Figure 3. 
\begin{figure}[htb]
\caption{The accelerometer is designed to detect the motion of the frame 
{\bf A.F.} defined by the accelerometer cage with respect to an inertial
frame {\bf I.F}. Any acceleration, seen as an inertial force $F$ acting on
the proof mass $M$, is detected by a capacitive sensor $CS$. The signal of
this sensor is used for the force detection $D$ as well as for keeping the
mass centered with respect to the cage through a servo-control loop $SL$.}
\label{fig3}
\end{figure}

The instrument is designed to detect the acceleration of the accelerometer
cage due to any external force. To this aim the relative motion of the proof
mass $M$ with respect to the frame defined by the cage is measured by the
capacitive sensor. An important characteristics of the mass is its free
mechanical impedance determined by a restoring force to the center of the
cage with a stiffness $K$ and a viscous damping with a coefficient $H_{m}$.
Depending on the physical origin of these effects, $K$ and $H_{m}$ may be
frequency dependent.

Dedicated to space applications, the accelerometer operation is based on the
electrostatic suspension of the proof mass in all spatial directions. Hence,
the mass is kept centered with respect to its cage through $3$ servo-control
loops demanded at least for stability (Earnshaw theorem). The acceleration
signal is in fact extracted from the knowledge of the electrostatic force
necessary to maintain the mass centered. In the real device, the control of
position and attitude is performed by six servo-control channels acting
separately. For simplicity, only one of the channels, corresponding to a
translation degree of freedom, is analyzed in this paper. 
\begin{figure}[htb]
\caption{Scheme of the capacitive sensor. The proof mass is placed between
two electrodes. The position dependent capacitances are polarized by an AC
sinewave source which induces a mean current at frequency $\protect\omega %
_{t}$ in the symmetrical mode. The mass displacement is read as the current
induced in the antisymmetric mode. An additional capacitance $C_{2}$ is
inserted to make the antisymmetric mode resonant with $\protect\omega _{t}$.
The electrical losses due to the quality factor of the transformer are
modeled as a resistance $R_{l}$ for the antisymmetric mode. The signal is
detected after an ideal operational amplifier with capacitive feedback $C_{f}
$ followed by a synchronous demodulation (not represented on this picture).
The impedance of the detection line plays the role of a further resistance $%
R_{r}$. The detected signal then feds the servo loop used to keep the mass
centered with respect to the cage.}
\label{fig4}
\end{figure}

As depicted on figure 4, the proof mass is placed between two symmetric
electrodes supported by the instrument cage which create two position
dependent capacitances. When the mass is centered in its cage, both
capacities are equal and the capacitance bridge is balanced. A displacement
of the mass creates an asymmetry of the bridge detected thanks to a
differential transformer and a pumping signal applied on the mass.
Conversely, voltages applied on these electrodes allow to exert
electrostatic forces on the mass. Capacitances are thus used for position
sensing as well as for generating the suspension force. Coupling between the
primary and secondary coils of the transformer being assumed ideal, the
transformer can be replaced by the equivalent circuit presented on Figure 5.
It is considered from now on that this transformation has been performed and
the circuit impedances redefined accordingly. 
\begin{figure}[htb]
\caption{Equivalent scheme of the capacitive sensor. The transformer of
figure 4 is replaced by the two inductances $L/2$ while the associated
losses are modeled as a resistance $R_{l}$. The other impedances are
modified accordingly.}
\label{fig5}
\end{figure}

The capacitances are polarized by an AC\ source of frequency $\omega _{t}$
which is chosen large enough for avoiding electrical $1/f$ noise and for
using low noise electronics. The sinewave source $E_{t}$ induces current at
frequency $\omega _{t}$ in the transformer symmetrical mode. In this static
and symmetric configuration, the current in the antisymmetric mode is zero
and the fluctuations of the two modes are uncoupled. Then a motion of the
proof mass at frequency $\Omega $ induces an asymmetry in the system and
creates sidebands on this electrical carrier $\omega _{t}$. The effect of
this asymmetry will be treated in a linear approximation with respect to the
deviations from the steady state equilibrium. The current induced in the
antisymmetric mode is thus proportional to the current in the symmetrical
mode and to the mass displacement. With this approximation, the fluctuations
of the symmetrical mode remain uncoupled to the antisymmetric mode and to
the mass motion. This is why the symmetric mode will be disregarded in the
following. In order to optimize the signal to noise ratio, an additional
capacitance $C_{2}$ is inserted which makes the antisymmetric mode resonant
with $\omega _{t}$. The electrical losses are mainly due to the quality
factor of the transformer and they are modeled by a resistance $R_{l}$ for
the antisymmetric mode.

The signal imprinted on the antisymmetric mode is detected after an ideal
operational amplifier with capacitive feedback (charge amplifier) followed
by a synchronous demodulation. This provides a low frequency voltage
proportional to the displacement of the mass. In a quantum network approach,
the signal is delivered by the capacitive sensor as the output field of a
detection line the impedance of which plays the role of a further resistance 
$R_{r}$. The description of the sensor is given in more detail in the next
section. This signal is used to feed the servo loop and keep the mass at its
equilibrium. Through the mass motion, it contains information on the
external forces acting on the mass. The noise added by the measurement
device to the measured observable is evaluated in the next sections, by
considering input fluctuations coming from all noise lines in the quantum
network model of Figure 6. 
\begin{figure}[htb]
\caption{Description of the accelerometer as a quantum network performing
input output tranformations on a number of lines. $m$ is the mechanical line
describing mechanical fluctuations as well as the measured signal, that is
the external force $F_{ext}$. $r_{1}$ is the detection line. $\protect\alpha 
$ labels the other lines $a_{1},a_{2},b_{1},b_{2},r_{1},r_{2},l_{1},l_{2}$
which contribute to noise. }
\label{fig6}
\end{figure}

It is in fact impossible to reach a stable equilibrium with a passive
electrostatic configuration. This is why the mass is actively maintained at
its equilibrium position by the generated electrostatic forces tailored
through the servo-control loop. The feedback control includes a proportional
and a derivation term. The generated electrostatic force proportional to the
measured mass displacement defines the servo-loop stiffness and, more or
less, the measurement bandwidth of the accelerometer. The force proportional
to the mass velocity introduces a motion damping to the benefit of the
control loop stability. This technique of active friction is equivalent to
an effective damping with reduced fluctuations in comparison to those
necessarily associated with a passive mechanical damping. This is why it is
called a cold damping technique. It will turn out that the added
fluctuations may even be smaller than the fluctuations associated with the
residual mechanical friction although the latter is much less efficient than
the active friction.

\section{The capacitive sensor}

In this section the capacitive sensor is analyzed in the absence of servo
control loop, with the equivalent electrical circuit of figure 5.

For a mass motion to be detected at frequency $\Omega $, the signal is
transposed by the electromechanical transducer to sidebands $\omega =\pm
\omega _{t}+\Omega $ of the carrier frequency $\omega _{t}$. The electrical
quadratures are defined as in equations (\ref{defquadrat}) and they are
dealt with separately so that the transducer appears as a three port
network. The first port is a mechanical one and corresponds to the velocity $%
V_{fr}$ of the free running proof mass and the force $F$ exerted on it. The
two other ports are electrical ones with the voltages $U_{t,n}$ and currents 
$I_{t,n}$ of the two quadratures $n=1,2$. The three port network is
described by an electromechanical impedance matrix 
\begin{eqnarray}
F &=& \left(\frac{iK}{\Omega }-iM\Omega \right) V_{fr} + \varkappa _{t}Z_{t}
I_{t1}  \nonumber \\
U_{t1} &=& Z_{t} I_{t1}  \nonumber \\
U_{t2} &=& 2i\varkappa _{t}Z_{t}\frac{\omega _{t}}{\Omega } V_{fr} + Z_{t}
I_{t2}  \nonumber \\
Z_{t} &=&-\frac{1}{2i\Omega C_{t}}  \label{Ztransduct}
\end{eqnarray}
$\frac{iK}{\Omega }-iM\Omega $ is the reactive part of the mechanical
impedance of the proof mass expressed in terms of mass $M$ and stiffness $K$%
. $Z_{t}$ is the electrical impedance evaluated at both frequencies $\pm
\omega _{t}+\Omega $ for a resonant circuit tuned at the polarization
frequency $\omega _{t}$. $\varkappa _{t}$ is an electromechanical coupling
constant proportional to the amplitude of the field created by the sinewave
electrical source applied to the mass. This impedance matrix shows that the
mechanical motion can be detected through the electrical quadrature $2$
whereas it is unaffected by the fluctuations coming through this port.
Meanwhile the mechanical motion is affected by the input fluctuations of the
electrical quadrature $1$. These features, typical of a quantum non
demolition coupling between electrical and mechanical elements, is discussed
in more detail in \cite{Francesca98}.

Fluctuations associated with losses are taken into account as the input
fields $l^{{\rm in}}$ coming to the transducer through the electrical line
of impedance $R_{l}$ as in equation (\ref{resistance}) and as the input
fields $m^{{\rm in}}$ coming through the mechanical line of impedance $H_{m}$
as in equation (\ref{defLangevin}). The external force $F_{ext}$ to be
detected comes as a mean field superimposed to the fluctuations $m^{{\rm in}%
} $ so that the equation of motion of the free running mass may be written 
\begin{eqnarray}
\Xi _{m}V_{fr} &=&F_{ext}-\varkappa _{t}Z_{t}I_{t1}-\sqrt{2\hbar \left|
\Omega \right| H_{m}}m^{{\rm in}}  \nonumber \\
\Xi _{m} &=&H_{m}-iM\Omega +\frac{iK}{\Omega }  \label{massmotion}
\end{eqnarray}
$\Xi _{m}$ is the full mechanical impedance of the proof mass in its free
running regime, now including not only the reactive part but also the
damping coefficient $H_{m}$.

The voltage and currents fluctuations associated with the amplifier have
then to be considered. In the configuration studied here, equations (\ref
{amplifier}) are replaced by 
\begin{eqnarray}
U_{l}\left[ \omega \right] &=&U_{t}\left[ \omega \right] =U_{r}\left[ \omega %
\right] +Z_{f}I_{f}\left[ \omega \right]  \nonumber \\
&=&\sqrt{2\hbar \left| \omega \right| R_{a}}\left( a^{{\rm in}}\left[ \omega %
\right] -b^{{\rm in}}\left[ -\omega \right] \right)  \nonumber \\
I_{l}\left[ \omega \right] +I_{f}\left[ \omega \right] +I_{t}\left[ \omega %
\right] &=&\sqrt{\frac{2\hbar \left| \omega \right| }{R_{a}}}\left( a^{{\rm %
in}}\left[ \omega \right] +b^{{\rm in}}\left[ -\omega \right] \right) 
\nonumber \\
Z_{f} &=&\frac{1}{-i\omega _{t}C_{f}}  \label{elec}
\end{eqnarray}
with $C_{f}$ the capacitor in the feedback loop of the amplifier. To
complete the set of equations associated with the electromechanical
transducer, the detected signal is the output field $r^{{\rm out}}$ which
comes out from the line $r$ of impedance $R_{r}$ and is therefore related to
the voltage $U_{r}$ as in equation (\ref{resistance}).

Equations (\ref{Ztransduct}-\ref{elec}) may be solved to obtain the output
field as well as the mass velocity. The latter quantity is expressed in
terms of the input fields ($\alpha $ labels the input noise lines $%
m,a_{1},a_{2},b_{1},b_{2},r_{1},r_{2},l_{1},l_{2}$; see Figure 6) 
\begin{eqnarray}
\Xi _{m}V_{fr} &=&F_{ext}+\sum_{\alpha }\lambda _{\alpha }\alpha ^{{\rm in}}
\nonumber \\
\lambda _{m} &=&-\sqrt{2\hbar \left| \Omega \right| H_{m}}  \nonumber \\
\lambda _{a_{1}} &=&-\lambda _{b_{1}}=-\sqrt{2\hbar \omega _{t}R_{a}}%
\varkappa _{t}  \nonumber \\
\lambda _{a_{2}} &=&\lambda _{b_{2}}=\lambda _{r_{1}}=\lambda
_{r_{2}}=\lambda _{l_{1}}=\lambda _{l_{2}}=0  \label{Vfree}
\end{eqnarray}
The velocity of the proof mass coupled to the electromechanical transducer
thus appears as a linear combination of the external force $F_{ext}$ to be
measured and of input fields in the noise lines associated either with
dissipative elements or with active ones. A number of coefficients $\lambda
_{\alpha }$ are null as a consequence of our symplifying assumptions, in
particular the assumption of the ideal operational amplifier. There remain
only two contributions to be discussed. The first corresponds to the
Langevin force fluctuations associated to the mechanical damping and
proportional to fields $m^{{\rm in}}.$ The second one comes from the voltage
noise at the input of the amplifier which is transformed to a back action
force exerted on the mass by the capacitive transducer. Accordingly, the
noise spectrum characterizing the velocity fluctuations is the sum of two
contributions which depend on the effective temperatures $\Theta _{m}$ and $%
\Theta _{a}$ associated respectively with the mechanical and the
amplification noise through (\ref{effectempM}) and (\ref{effectemp}) 
\begin{eqnarray}
\left| \Xi _{m}\right| ^{2}\sigma _{V_{fr}V_{fr}} &=&\sum_{\alpha }\left|
\lambda _{\alpha }\right| ^{2}\sigma _{\alpha \alpha }^{{\rm in}}  \nonumber
\\
&=&2H_{m}k_{B}\Theta _{m}+8R_{a}\varkappa _{t}^{2}k_{B}\Theta _{a}
\label{sigmaVVfree}
\end{eqnarray}

The output signal $r_{1}^{{\rm out}}$ is then evaluated by solving the same
equations (\ref{Ztransduct}-\ref{elec}). As the velocity, it is a linear
combination of the external force $F_{ext}$ and of input fields in the
various noise lines. When the expression of $r_{1}^{{\rm out}}$ is
normalized so that the coefficient of proportionality appearing in front of $%
F_{ext}$ is reduced to unity, the force estimator $\widehat{F}_{ext}$ is
just the sum of this external force to be measured and of the equivalent
input force noise 
\begin{eqnarray}
\widehat{F}_{ext} &=&\sqrt{\frac{\hbar R_{r}}{2\omega _{t}}}\frac{\Omega \Xi
_{m}}{2\varkappa _{t}Z_{f}}r_{1}^{{\rm out}}  \nonumber \\
&=&F_{ext}+\sum_{\alpha }\mu _{\alpha }\alpha ^{{\rm in}}  \label{estimfree}
\end{eqnarray}
The coefficients $\mu _{\alpha }$ are found to be 
\begin{eqnarray}
\mu _{m} &=&-\sqrt{2\hbar \left| \Omega \right| H_{m}}  \nonumber \\
\mu _{l_{2}} &=&-\frac{i\Omega \sqrt{\hbar }}{\sqrt{2R_{l}\omega _{t}}%
\varkappa _{t}}\Xi _{m}\qquad \mu _{l_{1}}=0  \nonumber \\
\mu _{r_{1}} &=&-\frac{\Omega \sqrt{\hbar R_{r}}}{2\sqrt{2\omega _{t}}%
Z_{f}\varkappa _{t}}\Xi _{m}\qquad \mu _{r_{2}}=0  \nonumber \\
\mu _{a_{1}} &=&-\mu _{b_{1}}=\sqrt{2\hbar R_{a}\omega _{t}}\left(
-\varkappa _{t}+\frac{\Omega }{2\varkappa _{t}\omega _{t}Z_{f}}\Xi
_{m}\right)  \nonumber \\
\mu _{a_{2}} &=&-\frac{i\Omega \sqrt{\hbar R_{a}}}{\sqrt{2}\varkappa _{t}%
\sqrt{\omega _{t}}}\Xi _{m}\left( \frac{1}{R_{a}}-\frac{1}{R_{l}}-\frac{1}{%
Z_{t}}\right)  \nonumber \\
\mu _{b_{2}} &=&-\frac{i\Omega \sqrt{\hbar R_{a}}}{\sqrt{2}\varkappa _{t}%
\sqrt{\omega _{t}}}\Xi _{m}\left( \frac{1}{R_{a}}+\frac{1}{R_{l}}+\frac{1}{%
Z_{t}}\right)  \label{mufree}
\end{eqnarray}
The comparison of equations (\ref{Vfree}) and (\ref{mufree}) shows that all
the terms $\lambda _{\alpha }$ of the expression (\ref{Vfree}) are found
present in (\ref{mufree}). The additional terms are interpreted as the
electrical noise due to the detection process. The force estimator (\ref
{estimfree}) can then be rewritten as 
\begin{equation}
\widehat{F}_{ext}=\Xi _{m}\left( V_{fr}+V_{se}\right)  \label{Velec}
\end{equation}
where $\Xi _{m}V_{fr}$ is given by (\ref{Vfree}) while $\Xi _{m}V_{se}$
collects all the other terms appearing in (\ref{mufree}). Because of the
normalization (\ref{estimfree}), these terms can be identified as those
which are proportional to $\Xi _{m}$. Physically, they represent the sensing
error. They involve amplifier current and voltage noise as well as Nyquist
noise associated to the loss and detection electrical lines. Since the
amplifier voltage noise is present in both contributions $\Xi _{m}V_{fr}$
and $\Xi _{m}V_{se}$, it follows that these two contributions are not
independent sources of noise.

The sensor noise spectrum $\Sigma _{FF}$, {\it i.e.} the noise associated
with fluctuations of $\left( \widehat{F}_{ext}-F_{ext}\right) $, is now
expressed as 
\begin{equation}
\Sigma _{FF}=\sum_{\alpha }\left| \mu _{\alpha }\right| ^{2}\sigma _{\alpha
\alpha }^{{\rm in}}  \label{addednoise}
\end{equation}
As a consequence of the preceding discussion, this added noise spectrum can
be written 
\begin{eqnarray}
\Sigma _{FF} &=&\left| \Xi _{m}\right| ^{2}\left( \sigma
_{V_{fr}V_{fr}}+\sigma _{V_{se}V_{se}}+\sigma _{V_{fr}V_{se}}\right) 
\nonumber \\
\sigma _{V_{fr}V_{fr}} &=&\frac{2H_{m}k_{B}\Theta _{m}}{\left| \Xi
_{m}\right| ^{2}}+\frac{8R_{a}\varkappa _{t}^{2}k_{B}\Theta _{a}}{\left| \Xi
_{m}\right| ^{2}}  \nonumber \\
\sigma _{V_{se}V_{se}} &=&\frac{\Omega ^{2}}{\omega _{t}^{2}\varkappa
_{t}^{2}}\left( \frac{1}{2R_{l}}k_{B}\Theta _{l}+\frac{R_{r}}{8\left|
Z_{f}\right| ^{2}}k_{B}\Theta _{r}\right.  \nonumber \\
&&\left. +R_{a}\left( \frac{1}{\left| Z_{f}\right| ^{2}}+\frac{1}{R_{a}^{2}}%
+\left| \frac{1}{R_{l}}+\frac{1}{Z_{t}}\right| ^{2}\right) k_{B}\Theta
_{a}\right)  \nonumber \\
\sigma _{V_{fr}V_{se}} &=&4R_{a}C_{f}\frac{K-M\Omega ^{2}}{\left| \Xi
_{m}\right| ^{2}}k_{B}\Theta _{a}  \label{addedfree}
\end{eqnarray}
The first two terms correspond to the noise spectrum of the velocity, the
terms proportional to the factor $\left| \Xi _{m}\right| ^{2}$ represent the
noise added by electrical detection. Finally the last line describes the
result of the interference between these two contributions.

\section{The cold damped accelerometer}

The cold damped accelerometer consists in the sensor studied in the
preceding section and the feedback loop used to generate the voltages
applied on the electrodes to control the mass motion.

The motion is measured through the sensor signal $r_{1}^{{\rm out}}$
previously described after a synchronous demodulation. The feedback force
applied for controlling the motion of the mass is obtained through a low
frequency\ amplifier. The set of equations describing the complete
accelerometer is the same as in the previous section (\ref{Ztransduct}-\ref
{elec}) except for the equation of the proof mass motion (\ref{massmotion})
which is now read as 
\begin{eqnarray}
\Xi _{m}V_{cd}&=&F_{ext}-\varkappa _{t}Z_{t}I_{t1}  \nonumber \\
&&-\sqrt{2\hbar \left| \Omega \right| H_{m}}m^{{\rm in}}-G_{s}r_{1}^{{\rm out%
}}+F_{s}^{{\rm in}}  \label{massmotionservo}
\end{eqnarray}
$V_{cd}$ now denotes the velocity of the proof mass in presence of the cold
damping. The term $G_{s}r_{1}^{{\rm out}}$ represents the feedback action on
the mass with the whole gain of the servo loop denoted $G_{s}$. The
impedance of the detection line $r$ is assumed to be small $R_{r}\ll \left|
Z_{f}\right| $ so that its contribution is negligeable. It is therefore
equivalent to add a feedback proportional to $r^{out}$ or proportional to
the output voltage of the amplifier $U_{r}$. $F_{s}^{{\rm in}}$ are the
force fluctuations due to the active and passive elements used to generate
the servo control force.

The solution of these equations yields the velocity of the cold damped mass 
\begin{eqnarray}
&&\left( \Xi _{m}+\Xi _{me}\right) V_{cd}=F_{ext}+\sum_{\alpha }\lambda
_{\alpha }\alpha ^{{\rm in}}+\sum_{\beta }\lambda _{\beta }\beta ^{{\rm in}}
\nonumber \\
&&\Xi _{me}=H_{me}+\frac{iK_{me}}{\Omega }=-\sqrt{\frac{2\omega _{t}}{\hbar
R_{r}}}\frac{2\varkappa _{t}Z_{f}}{\Omega }G_{s}  \label{vitessecold}
\end{eqnarray}
The servo loop produces an effective mechanical impedance $\Xi _{me}$
written as the sum of a damping term $H_{me}$ and a restoring force of
stiffness $K_{me}$, both parameters being frequency dependent. In particular 
$K_{me}$ can include the effect of an integrator term in the feedback
corrector. This term ensures the motionlessness of the mass at very low
frequencies to the benefit of the instrument accuracy. The noise terms $%
\lambda _{\alpha }\alpha ^{{\rm in}}$ represent the fluctuations due to the
input fields $\alpha ^{{\rm in}}$ as in the previous section. In addition,
there are noise terms $\lambda _{\beta }\beta ^{{\rm in}}$ added by the
active and passive elements in the servo loop.

Let us consider now the actual instrument case where the effective
mechanical impedance $\Xi _{me}$ is much larger than the free mass impedance 
$\Xi _{m}$ 
\begin{eqnarray}
H_{me} &\gg &H_{m}  \nonumber \\
K_{me} &\gg &\left| K-M\Omega ^{2}\right| 
\end{eqnarray}
These conditions are fully compatible with the stability of the feedback as
evaluated in the design and demonstrated with the real instruments 
\cite{Bernard91,Touboul92,Willemenot97}. In the equations of motion written
previously, this case corresponds to the limit of an infinite loop gain 
\begin{equation}
G_{s}\rightarrow \infty 
\end{equation}
Then, the noise terms $\lambda _{\beta }\beta ^{{\rm in}}$ coming from the
servo loop scale as $\sqrt{G_{s}}$. Hence their effect on velocity scales as 
\begin{equation}
\frac{\lambda _{\beta }}{G_{s}}\varpropto \frac{1}{\sqrt{G_{s}}}\rightarrow 0
\end{equation}
so that they may be forgotten in (\ref{vitessecold}). This only means that,
as well known, the dominant noise sources are those associated with the
first amplification stage, here the terms $\lambda _{\alpha }\alpha ^{{\rm in%
}}.\;$

The velocity (\ref{vitessecold}) stabilized by the feedback loop is now read
as 
\begin{eqnarray}
V_{cd} &=&-\sqrt{\frac{\hbar R_{r}}{2\omega _{t}}}\frac{\Omega }{2\varkappa
_{t}Z_{f}}\sum_{\alpha }\frac{\lambda _{\alpha }}{G_{s}}\alpha ^{{\rm in}} 
\nonumber \\
\frac{\lambda _{m}}{G_{s}} &=&0  \nonumber \\
\frac{\lambda _{l_{2}}}{G_{s}} &=&-\frac{2iZ_{f}}{\sqrt{R_{l}R_{r}}}\qquad 
\frac{\lambda _{l_{1}}}{G_{s}}=0  \nonumber \\
\frac{\lambda _{r_{1}}}{G_{s}} &=&-1\qquad \frac{\lambda _{r_{2}}}{G_{s}}=0 
\nonumber \\
\frac{\lambda _{a_{1}}}{G_{s}} &=&-\frac{\lambda _{b_{1}}}{G_{s}}=2\sqrt{%
\frac{R_{a}}{R_{r}}}  \nonumber \\
\frac{\lambda _{b_{1}}}{G_{s}} &=&-2iZ_{f}\sqrt{\frac{R_{a}}{R_{r}}}\left( 
\frac{1}{R_{a}}-\frac{1}{R_{l}}-\frac{1}{Z_{t}}\right)  \nonumber \\
\frac{\lambda _{b_{2}}}{G_{s}} &=&-2iZ_{f}\sqrt{\frac{R_{a}}{R_{r}}}\left( 
\frac{1}{R_{a}}+\frac{1}{R_{l}}+\frac{1}{Z_{t}}\right)  \label{lambdacold}
\end{eqnarray}
Since the servo loop efficiently maintains the mass at its equilibrium
position, the velocity is no longer affected by the external force $F_{ext}$%
. However the sensitivity to external force is still present in the
correction signal which will be discussed later on. The residual motion of
the mass is described by the various noise terms $\frac{\lambda _{\alpha }}{%
G_{s}}\alpha ^{{\rm in}}$. The values of these coefficients are easily
interpreted through a comparison with the force estimator (\ref{estimfree})
evaluated in the preceding section for the capacitive sensor. The cold
damped motion of the proof mass is indeed described by the simple equation 
\begin{equation}
V_{cd}=-V_{se}  \label{Vcold}
\end{equation}
where $V_{se}$ is the difference between the real velocity of the mass and
the velocity measured by the sensor. This means that the servo loop
efficiently corrects the motion of the mass except for the sensing error $%
V_{se}$.

With the same set of equations (\ref{Ztransduct}-\ref{elec} with \ref
{massmotionservo} replacing \ref{massmotion}), the output field $r_{1}^{{\rm %
out}}$ is evaluated and exploited as a measurement of the external force. As
in the previous section, this output field is normalized so that the force
estimator $\widehat{F}_{ext}$ appears as the sum of the real force and of an
equivalent force noise 
\begin{eqnarray}
\widehat{F}_{ext} &=&\sqrt{\frac{\hbar R_{r}}{2\omega _{t}}}\frac{\Omega \Xi
_{me}}{2\varkappa _{t}Z_{f}}r_{1}^{{\rm out}}  \nonumber \\
&=&F_{ext}+\sum_{\alpha }\mu _{\alpha }\alpha ^{{\rm in}}  \label{estimcold}
\end{eqnarray}
This expression is similar to the estimator (\ref{estimfree}) evaluated for
the free mass although the free impedance $\Xi _{m}$ has been replaced by
the effective impedance $\Xi _{me}$.

A quite remarkable result is then obtained. In the limit of the infinite
loop gain and with the same approximations as above, the expressions of the
coefficients $\mu _{\alpha }$ are exactly the same as those (\ref{mufree})
corresponding to the open loop case. The expression of the force estimator $%
\widehat{F}_{ext}$ is the same as in the free case while the expression of
the velocity is quite different. The actual motion of the mass is indeed
independent of the external perturbations in the servo control case with the
velocity determined by the sensor noise (\ref{Vcold}).

It is in fact possible to reexpress the force estimator (\ref{estimcold}) as
the sum of two terms 
\begin{equation}
\widehat{F}_{ext}=\Xi _{m}\left( V_{fr}+V_{se}\right) =\Xi _{m}\left(
V_{fr}-V_{cd}\right)
\end{equation}
The first term is exactly the same as the actual motion (\ref{Vfree}) of the
free running mass. It is the sum of the external force $F_{ext}$ and of the
force fluctuations exerted on the mass in the absence of servo control,
namely the mechanical Langevin force and the back action force due to the
sensor. The second term is the actual velocity (\ref{Vcold}) of the mass
that is also the already discussed sensor error. Once again, these two terms
are correlated since both depend on the same amplifier voltage noise. The
expression of the noise spectrum $\Sigma _{FF}$ is not reproduced here since
it is exactly the same (\ref{addedfree}) as in the open loop case.

\section{Discussion}

The results obtained in the two previous sections allow to evaluate the
performance of the cold damping technique for a wide range of experimental
parameters and for all temperatures. In this concluding section, we want to
discuss these results by focussing our attention on the present
state-of-the-art instrument as well as on ultimate sensitivity limits which
can be reached in the future with such an accelerometer.

The noise spectrum for the velocity of the proof mass in its free running
regime may be rewritten 
\begin{equation}
H_{m}\sigma _{V_{fr}V_{fr}}=\frac{2}{1+\Delta ^{2}}\left( k_{B}\Theta _{m}+4%
\frac{R_{a}}{R_{m}}k_{B}\Theta _{a}\right)  \label{VfVf}
\end{equation}
The parameter $\Delta $ measures the reactive impedance of the free mass as
compared to the dissipative one 
\begin{equation}
\frac{K}{\Omega }-M\Omega =H_{m}\Delta  \label{defDelta}
\end{equation}
The electrical resistance $R_{m}$ allows to express the mechanical damping
coefficient $H_{m}$ through the conversion relation 
\begin{equation}
R_{m}=\frac{H_{m}}{\varkappa _{t}^{2}}  \label{Rm}
\end{equation}
With this definition, the ratio $\frac{R_{a}}{R_{m}}$ allows to compare the
electrical and mechanical noises in (\ref{VfVf}).

The noise spectrum $\sigma _{V_{se}V_{se}}$ for the sensing error $V_{se}$,
which is also the noise $\sigma _{V_{cd}V_{cd}}$ for the velocity $V_{cd}$
of the proof mass in the cold damped regime, is expressed in a similar form 
\begin{eqnarray}
H_{m}&&\sigma _{V_{se}V_{se}} = \frac{\Omega ^{2}}{\omega _{t}^{2}}\left( 
\frac{R_{m}}{R_{l}}k_{B}\Theta _{l}+\frac{R_{m}R_{r}}{4\left| Z_{f}\right|
^{2}}k_{B}\Theta _{r}\right.  \nonumber \\
&&\left. +2R_{a} R_{m} k_{B} \Theta _{a} \left( \frac{1}{\left| Z_{f}\right|
^{2}}+\frac{1}{R_{a}^{2}}+\left| \frac{1}{R_{l}}+\frac{1}{Z_{t}}\right|
^{2}\right) \right)  \label{VeVe}
\end{eqnarray}
The sensing error is minimized by diminishing the fluctuations coming from
the electrical noise lines, that is when the transducer impedance $Z_{t}$,
the feedback impedance $Z_{f}$ and the loss impedance $R_{l}$ are chosen
high enough. The transposition ratio $\frac{\Omega ^{2}}{\omega _{t}^{2}}$
appears as a common factor which greatly helps in keeping this error low.

The two contributions (\ref{VfVf}) and (\ref{VeVe}) are added in the whole
added noise spectrum $\Sigma _{FF}$ together with a third term $\sigma
_{V_{fr}V_{cd}}$ 
\begin{equation}
H_{m}\sigma _{V_{fr}V_{se}}=4\frac{R_{a}}{\left| Z_{f}\right| }\frac{\Omega 
}{\omega _{t}}\frac{\Delta }{1+\Delta ^{2}}k_{B}\Theta _{a}  \label{VfVe}
\end{equation}
This term is also reduced when the feedback impedance $Z_{f}$ is large.

Let us evaluate the whole noise spectrum $\Sigma _{FF}$ for the specific
case of the instrument proposed for the $\mu $SCOPE space mission devoted to
the test of the equivalence principle. The parameters have the following
values 
\begin{eqnarray}
M=0.27 {\rm \ kg} &\qquad& H_{m} = 1.3 \times 10^{-5} {\rm \ kg \ s}^{-1} 
\nonumber \\
K=4 \times 10^{-6} {\rm \ N \ m}^{-1} &\qquad& \Delta \simeq 100  \nonumber
\\
\frac \Omega {2\pi} \simeq 5 \times 10^{-4} {\rm \ Hz} &\qquad& \frac {%
\omega_t} {2\pi} \simeq 10^{5} {\rm \ Hz}  \nonumber \\
\varkappa _{t}=10^{-7} {\rm \ C \ m}^{-1} &\qquad& R_{l}=2.5 \times 10^{5} \
\Omega  \nonumber \\
R_{m}=1.3 \times 10^{9} \ \Omega &\qquad& \Theta _{m}=300 {\rm \ K} 
\nonumber \\
\left| Z_{f}\right| =1.6\times 10^{5} \ \Omega &\qquad& \left| Z_{t}\right|
=10^{14} \ \Omega  \nonumber \\
R_{a}=0.15 \times 10^6 \ \Omega &\qquad& \Theta _{a}=1.5 {\rm \ K}
\label{parameters}
\end{eqnarray}

In these conditions, the added noise spectrum is dominated by the mechanical
Langevin forces 
\begin{eqnarray}
\Sigma _{FF} &=&2H_{m}k_{B}\Theta _{m}  \nonumber \\
&=&1.1 \times 10^{-25} \left({\rm kg \ m \ s^{-2}}\right) ^{2}/ {\rm Hz}
\end{eqnarray}
This corresponds to a sensitivity in acceleration 
\begin{equation}
\frac{\sqrt{\Sigma _{FF}}}{M}=1.2 \times 10^{-12} {\rm \ m \ s^{-2}} / \sqrt{%
{\rm Hz}}
\end{equation}
Taking into account the integration time of the experiment, this is
consistent with the expected instrument performance corresponding to a test
accuracy of $10^{-15}$.

In the present state-of-the-art instrument, the sensitivity is thus limited
by the residual mechanical Langevin forces. The latter are due to the
damping processes in the gold wire used to keep the proof mass at zero
voltage \cite{Willemenot97}. With such a configuration, the detection noise
is not a limiting factor. This is a remarkable result in a situation where
the effective damping induced through the servo loop is much more efficient
than the passive mechanical damping. This confirms the considerable interest
of the cold damping technique for high sensitivity measurement devices.

Future fundamental physics missions in space will require even better
sensitivities. To this aim, the wire will be removed and the charge of the
test mass will be controlled by other means, for example UV photoemission.
The mechanical Langevin noise will no longer be a limitation so that the
analysis of the ultimate detection noise will become crucial for the
optimization of the instrument performance. This also means that the
electromechanical design configuration will have to be reoptimized taking
into account the various noise sources associated with detection.

In order to evaluate these added noise sources we consider the whole noise
spectrum obtained by taking into account the spectra (\ref{VfVf}), (\ref
{VeVe}) and (\ref{VfVe}) 
\begin{eqnarray}
\Sigma _{FF} &=& H_{m} \left(1+\Delta ^2 \right) \left(\sigma _{V_{fr}
V_{fr}} + \sigma _{V_{se} V_{se}} + \sigma _{V_{fr} V_{se}} \right)
\end{eqnarray}
This spectrum contains terms scaling as $R_{m}$ as well as terms scaling as $%
\frac{1}{R_{m}}$. Hence, there exists an optimum value for $R_{m}$ when the
other parameters as fixed. In the same way, it includes terms scaling as $%
R_{a}$ and as $\frac{1}{R_{a}}$ so that there exists an optimum value for $%
R_{a}$. In contrast, the noise is always lowered by reducing the electrical
losses with large values for the impedances $Z_{t}$, $Z_{f}$ and $R_{l}$ and
low values for $R_{r}$.

In these limits, the added noise spectrum $\Sigma _{FF}$ takes a simple form 
\begin{eqnarray}
\Sigma _{FF} &=&2H_{m}k_{B}\Theta _{m}+8H_{m}\frac{R_{a}}{R_{m}}k_{B}\Theta
_{a}  \nonumber \\
&&+2H_{m}\left( 1+\Delta ^{2}\right) \frac{\Omega ^{2}}{\omega _{t}^{2}}%
\frac{R_{m}}{R_{a}}k_{B}\Theta _{a}
\end{eqnarray}
This final result is optimized by matching the values of the impedances $%
R_{a}$ and $R_{m}$ so that 
\begin{eqnarray}
&&\left( \frac{R_{a}}{R_{m}}\right) ^{{\rm opt}}=\frac{\sqrt{1+\Delta ^{2}}}{%
2}\frac{\left| \Omega \right| }{\omega _{t}}  \nonumber \\
&&\Sigma _{FF}^{{\rm opt}}=2H_{m}k_{B}\Theta _{m}+8H_{m}\sqrt{1+\Delta ^{2}}%
\frac{\left| \Omega \right| }{\omega _{t}}k_{B}\Theta _{a}
\label{finalresult}
\end{eqnarray}
This is the sum of the already discussed limit associated with mechanical
Langevin fluctuations and of a second term which represents the ultimate
detection noise. The first contribution dominates the second one for the
present state-of-the-art instrument but this will no longer be the case for
future instruments designed for better performance tests of the equivalence
principle. For such instruments, equation (\ref{finalresult}) shows that the
sensitivity may be largely improved.

{\bf Acknowledgements}

Thanks are due to Alain Bernard, Vincent Josselin and Eric Willemenot for
helpful discussions.

\end{document}